\documentclass[runningheads]{svmult}

\usepackage{makeidx}   
\usepackage{graphicx}  
\usepackage{subeqnar}  
\usepackage{multicol}  
\usepackage{physmubb}  
\makeindex             

\begin{document}

\parskip=0.6\baselineskip
\parindent=0pt
\title*{
{\small 
{\hfill hep-ex/0211009, Fermilab-Conf-02/275-E}\\} 
{\ }\\
Summary and Highlights of the\\
14th Topical Conference on Hadron 
Collider Physics (HCP2002)}
\titlerunning{Summary and Highlights of HCP2002}
%
\author{
{\bf{Karlsruhe, Germany, September 29 -- October 4, 2002}}\\
\bigskip
\bigskip
John Womersley\\
{\em Fermi National Accelerator Laboratory, Batavia, IL 60510, U.S.A.\\
\bigskip
November 4, 2002}
}
\authorrunning{John Womersley}
%
%

\maketitle              

\section{Introduction}
First of all, I would like to thank the scientific committee, the conference
organizers, the University of Karlsruhe and the Institute for Experimental
Nuclear Physics, all of the speakers, and the conference secretariat, for 
making this an extremely well-organized and uniformly high-quality meeting.
I would also like to thank all of the speakers who provided me with
material for my talk before and during the conference.

There is obviously no point in these proceedings in attempting to 
repeat all of the material from the individual contributions; by
definition, these are all available earlier in this volume. 
In the written version, therefore, I will try to give a high
level overview of the current state of hadron collider physics
and to highlight the connections between the many presentations
at this conference.

\section{Our Tools}
The tools of hadron collider physics
consist of accelerators, detectors, and computing infrastructure;
theoretical predictions
and simulation programs; our knowledge of the structure of the 
proton; and analysis techniques.

\subsection{Accelerators and Detectors}

At the Tevatron\cite{Harms}, 
we are not yet out of the woods, but gratifying progress in Run~II
was reported.   Record peak
($3.6\times 10^{31}{\rm cm}^{-2}{\rm s}^{-1}$)
and weekly ($6.7{\rm pb}^{-1}$) luminosities have been 
recorded in the last few weeks.  The complex is now exceeding its
performance in Run~I.  These improvements have come from specific modifications
to the complex, and from much hard work.   
The injectors are providing the necessary beam
for higher luminosities, and there is a fully resource-loaded
plan for the next year. The major issues are Tevatron 
transfer and acceleration efficiencies, emittance dilution, beam
lifetimes at 150\,GeV (before acceleration), 
and the role of long range beam-beam effects.
There is no silver bullet; rather, there are a large number of ten
to fifteen percent improvements to be made.

The CDF and D\O\ detectors are both working well and recording physics 
quality data.  D\O\cite{Buscher} showed results from tracking, 
calorimeter and muon detectors.  Improvements are still in store in
the trigger, including a silicon vertex detector currently
under construction.  CDF\cite{Chlebana} reported on detector and
trigger capabilities.  Track triggering uses the drift chamber at
level 1 and displaced tracks from silicon at level 2; this is a
major success and has already yielded some very impressive heavy 
flavor samples. A tau trigger\cite{Smith} is running and
a clear $W\rightarrow \tau \nu$ signal is seen. 

The HERA accelerator\cite{Niebuhr} has been substantially upgraded
since its last physics run in 2000. There are new interaction 
regions with $\sim 500\,$m of new accelerator, 58 new magnets,
and spin rotators.
The goal is to increase luminosity by a factor of roughly four and
to deliver about $1 {\rm fb}^{-1}$ to the experiments by 2006. 
Commissioning has been a painful process, thanks to synchrotron radiation
problems (better shielding is needed) and beam-gas interactions
(requiring improved vacuum).  The detectors have also been upgraded,
in pursuit of physics goals in QCD, proton structure and searches.

At the LHC \cite{Rolandi}, the challenges are those
of complexity and scale.  For the detectors, these take the form
of occupancy and radiation in the tracking detector, the need to build
huge detectors to measure high-$p_T$ muons, the need for excellent
electromagnetic energy resolution in the calorimeters 
(for $H \rightarrow\gamma\gamma$), and the challenges of triggering
and subsequent data processing.  The management,
logistical and assembly issues are also on a new scale.  The
scheduled accelerator start-up remains in 2007.  Dipole magnet
production is the critical path and it is still very early in
the process (40 of roughly 2000 dipoles are expected by the end
of 2002).  

\subsection{Luminosity Measurement}

Knowledge of the luminosity in a hadron collider requires counting
the rate of a reference process with a known cross section. 
Traditionally, the total inelastic cross section has been used.  
This allows instantaneous, real-time, bunch by bunch measurements.
The CDF and D\O\ detectors have installed new detectors to measure
the inelastic rate in Run II\cite{Klimenko}: scintillators in D\O,
and a novel {\v C}erenkov detector in CDF.  The inelastic cross section
can determine luminosity at the $\Delta{\cal L}\sim 3-5\%$ level, but
there are problems in knowing what cross section to use (in the
past, D\O\ used the World Average and CDF used their own measurement)
and what uncertainty to assign (these two cross sections were not really
compatible within their errors). These will be resolved for Run~II.

It has been proposed \cite{Dittmar,Giele} that greater 
precision could be obtained by using inclusive $W$ or $Z$ production
as the reference process.  This is a better known, calculable
cross section, but the acceptance needs to be modelled and the
calculated cross section depends on parton distributions (hence it
is important to understand the uncertainties on the latter).  
The feasibility of this idea has been demonstrated 
using D\O\ Run~I data, but  it has not yet been used by the
experiments; CDF and D\O\  will do this ``for real'' in Run~II.
Initially we expect  $\Delta{\cal L}\sim 3-5\%$ here, with
the hope of reducing the error to $\sim 1\%$ over time.

At the LHC\cite{Rijssenbeek}, ATLAS and CMS have set the
goal of $\Delta{\cal L}=2\%$.  They appear to be ``covering 
all the bases,'' with dedicated small angle detectors (TOTEM
in CMS, Roman Pots in ATLAS) and plans for special running
with a detuned $\beta^*$ to measure the reference process
at 14\,TeV; wider-angle forward detectors for real-time monitoring of
luminosity; and the potential to use physics processes like
$W$ and $Z$-production offline.

\subsection{Computing and Analysis}

Computing for data processing and analysis is a challenge
for modern experiments both because of the quantity of data and
the size and geography of the collaborations \cite{Kasemann}.  
There is a natural
synergy between the need to address these
challenges and current ideas about ``Grid''
computing.  Developments at the Tevatron are making something
like a Grid a reality (distributing data analysis using the
SAM system for both CDF and D\O), and the LHC will rely
much more on the full realization of distributed resources
and the tools to make them useful.

Given the low signal cross sections for many interesting
processes at hadron colliders, analysis of our data
clearly benefits from the use of advanced techniques
to separate signal from background.  Multivariate tools\cite{Koblitz} 
are now quite widely employed. 

\subsection{Simulation}

Event generators are essential tools for understanding hadronic
processes.  There is much effort towards improving the
showering event generators, such as Herwig and Pythia, that are 
very widely used in hadron collider physics\cite{Gieseke}
The parton shower can be corrected using the matrix
element to cover phase space better. There is work
to connect  parton showers with the leading order matrix elements
for many-body final states without
double-counting or dead regions of phase space -- 
very useful with the new generation of automatic 
matrix element calculators (see below).
There are also efforts to merge higher order
matrix elements with parton-showering event generators without
double-counting, perhaps the best known being the MC@NLO project. 

The Herwig event generator is currently being rewritten in C++,
and uses a common class library with Pythia7 (but not common 
physics functions).  A beta version for $e^+e^-$
is expected later in 2002, with the first release 
in 2003 and a full version for hadron colliders by 2004.

It would be very desirable to have a better modelling of
the underlying event in hadron-hadron collisions; this is
an important source of uncertainty on the jet energy scale
(and thus on the top mass).
To help, there are some very nice data from CDF \cite{Martinez}
on particle flow around and between jets.
Multiple parton scattering also needs to be understood:
what fraction of the $Wb\overline b$ background at the
LHC is due to a $W$ and a $b\overline b$ pair from
different parton-parton collisions?
 
\subsection {Proton Structure}

Our current knowledge of parton distributions is
dominated by deep inelastic scattering (DIS) data from HERA
\cite{Schnurbusch}.  Structure functions are
measured at the few {\%} 
level and cover six orders of magnitude in $x$ and $Q^2$.
Photon, $Z$ and $W^\pm$ exchange are probed.  With the
HERA~II upgrade, we can expect higher luminosity, more
$e^-$ data, 
and information from polarized beams.  In addition to
DIS, the latest CTEQ6 and MRST2001 fits use Tevatron jet
data to constrain the gluon distribution.

One can fit all the current data with NLO theory and
a reasonably consistent set of parton distributions\cite{Thorne}; 
the resulting $\chi^2$ is about 1.1 per degree
of freedom.  Most recent work has been directed at
understanding the uncertainties on these distributions
(inspired by, among other things, the controversy over
whether parton distributions were consistent with
the Tevatron high-$E_T$ jet cross section).
A variety of approaches is used by the different groups,
mainly differing in their treatment of systematic errors.
The uncertianties in the distributions are at the $1-5\%$ 
level except in odd regions (such as the gluon and $d$-quark
distributions at high-$x$).
Unfortunately, it appears that the uncertainties due to
varying the theoretical assumptions, $\alpha_s$, cuts on the
data and so on can be much greater than the experimental
errors.  As an example, if the parton distributions
are refitted after excluding DIS data below a
cut of $x=0.005$, the calculated Tevatron $W$ and Higgs 
cross sections shift by about three times the nominal 
uncertainty from the original parton distributions.
This probably points to inadequacies in the theoretical
predictions that are used to extract parton
distributions from measured DIS cross sections
(such as incomplete treatment of
higher order terms, low-$x$ or high-$x$
resummation, low-$Q^2$ or higher twist effects).

\subsection{Theoretical Progress}

There are various fronts on which theoretical progress at
hadron colliders is taking place.

Next-to-next-to-leading order (NNLO) calculations are
required to challenge the high statistics results from 
the Tevatron and HERA \cite{Glover,Gehrmann}.
This has been known for a while, but 
the bottleneck was calculation of the two-loop box
graph, a critical component of the NNLO jet cross section.
This has now been solved and there has been great progress
in the last couple of years; we can expect the first
NNLO parton level Monte Carlo generators in the next two
years, including $p\overline p\rightarrow {\rm jet}+X$.

Leading order simulations for up to 8 partons in the final
state are now available (e.g. QCD backgrounds to $t\overline t H$).
These and other matrix elements can now be automatically 
generated by the non-expert user with programs such as 
Madgraph, Comphep, Alpgen, etc.

Vector bosons plus jets are a critical background at the
Tevatron and LHC \cite{Giele} and need to be well
understood.  At Leading Order, $W/Z+$ any number of
jets is available.  At NLO, $W/Z + 2{\rm jets}$ is
handled by the {\tt mcfm} program, but there is a real need for
an NLO $W/Z + 3{\rm jets}$ parton level generator (this
is the dominant backgound to top).
The cross section calculations are reasonably stable at
NLO and there is good agreement with the data for inclusive
and one-jet final states; the data with up to four jets agrees
with the LO calculation after some scale tuning.  Vector
boson transverse momentum in the low-$p_T$ region
is not modelled well by these programs: non-perturbative
parameters are involved but can be extracted from the data
(using the $Z$ to model the $W$ for example).

\section{Our Physics}

In the opening presentation\cite{Willenbrock}, we heard five
ways in which hadron colliders can confront the Standard Model:
\begin{itemize}
\item The strong interaction
\item The CKM matrix
\item Electroweak measurements
\item The top quark
\item The Higgs boson
\end{itemize}
To these we should add
\begin{itemize}
\item Direct searches for new phenomena not part of the Standard Model.
\end{itemize}

\subsection {QCD}

No one doubts that QCD {\em is} the theory of the strong interaction
of quarks and gluons.  QCD is so central to the calculation
of signal and background processes at hadron colliders that we
need to make sure that we can have confidence in our ability
to make predictions in its framework.
We need to resolve some outstanding puzzles in the data, and ensure
we understand how to calculate the backgrounds to new physics. 

\subsection*{Jet Production}

Both CDF\cite{Martinez} and D\O\cite{Gallas} reported
inclusive jet $E_T$ distributions from Run~II.  While these are not yet 
fully corrected, already we see events out to $E_T \sim 400\,$GeV.  
CDF are making use of their new forward calorimetry to cover
the whole range $0.1<|\eta|< 3.0$.  With the full Run~II dataset
the inclusive distribution should extend out to $E_T > 600\,$GeV, 
allowing us to pin down the high-$E_T$ behaviour of the cross section
and providing tighter constraints on the gluon PDF. (Recall that
the gluon at high $x$ is one of the least well-constrained PDF's).

Another issue provoking much discussion was the choice of jet
algorithm \cite{Chekanov}.  The D\O\ Run~I data have shown that
cone and $k_\perp$ algorithms yield different cross sections
for collider jet data; this is expected and qualitatively agrees
with parton-level simulations.  Quantitatively, though, it is 
not yet fully clear
whether the difference is actually at the level expected and
whether showering and hadronization (not modelled
in a parton level Monte Carlo) can explain it.

At HERA, jet studies allow very precise tests of QCD predictions
\cite{Gonzalez}.
Studies are carried out in two regimes: photoproduction and deep
inelastic scattering (DIS).  There has been a big effort to reduce
the experimental uncertainties in tests of
perturbative QCD at HERA, and now the theoretical uncertainties
limit the precision attainaible in many analyses: higher orders
or resummed calculations are needed. Also, the present precision
in the predictions is not sufficient
to constrain the partonic content
of the photon. Both H1 and ZEUS have new, precise determinations
of $\alpha_s$ using DIS data; the uncertainties are comparable
to that in the current World Average.

\subsection*{Heavy flavour production}

At the Tevatron, Run~I left a lot of unanswered questions\cite{Bishai}.
The measured inclusive $B$ production cross section lies significantly
above the NLO QCD prediction, 
though the prediction can be made to agree better with resummation and
retuned $b\rightarrow B$ fragmentation (from LEP).  For charmonium, the 
observed cross section requires a large colour-octet component
which matches the $p_T$ distribution seen in data but then completely
fails to describe the $J/\psi$ polarization above $p_T ~ 10\,$GeV.
The CDF secondary vertex trigger in Run~II is working beautifully
and the resulting large charm and bottom samples will allow these
puzzles to be explored in much more detail.  D\O\cite{Bauer}
showed preliminary Run~II $J/\psi$ and muon+jet cross sections
as first steps in measuring the charmonium polarization (and
production process)
and the $b$-jet cross section.  

At HERA \cite{Meyer} there have been many new results on
charm and bottom production in 2002.  Almost all lie significantly
above NLO theory, though a ZEUS DIS measurement 
at the largest $Q^2$ is in agreement (note that
the same pattern 
is seen in the Tevatron $b$-jet cross section which
comes closer to QCD at the highest $p_T$).  There is no
hard evidence for (or against) a colour-octet contribution to
$J/\psi$ production at HERA. 

On the theoretical side\cite{Gehrmann,Kniehl}, 
there were suggestions that an incomplete treatment of fragmentation
may be part of the reason for the $B$-production ``excess.'' 
Also, Tevatron data on fragmentation have not been published since
the 1988-89 run and higher statistics would be useful.     
Given the uncertainty as to whether NLO theory is adequate
for heavy flavour, we really 
need NNLO calculations both for charmonium and $b$ production.

There is also the suggestion \cite{Palisok} that one could
test the charmonium production mechanism using vector boson
plus charmonium associated production.  

\subsection*{Direct Photons}

Isolated photon results from hadron collisions\cite{Lee}
are quite consistent
with NLO QCD at high $p_T$, but CDF, D\O\ and E706 data show
an excess at the low $p_T$ end of the spectrum.
One explanation proposed is that there is additional
transverse momentum (``$k_T$'') from soft gluon radiation.
One can model this using a few GeV of Gaussian transverse
smearing, resulting in a much improved match to the measured
cross sections. The amount of smearing needed increases smoothly from
about 1\,GeV at fixed target energies to 1.5\,GeV at HERA and 3.5\,GeV
at the Tevatron.  Resummation offers the hope of a more 
predictive calculation.  The fragmentation contribution 
to photon production cannot be neglected; again, LEP
results are used.  Also, at HERA, photon production may
indicate a need to review the present modelling of the
partonic structure of the photon.

\subsection*{QCD at the 1\,GeV scale}

In low-$x$ DIS at HERA \cite{Behnke}, DGLAP evolution of the 
structure function $F_2$ works all the way down
to about 2\,GeV$^2$.  Leading order DGLAP plus a resolved
photon describe the data well. Below this, a variety of
models are invoked: Regge, color dipoles, etc.  There is
no sign in the data of BFKL evolution or of saturation
effects.  
One place where the data seem to prefer BFKL -- in fact
the only place I can think of where this is the case --
is high-$t$ vector meson production at ZEUS\cite{Forshaw}.
A resolved photon interacting with a BFKL gluon ladder
describes the data well, and straightforward
two-gluon exchange fails to do so.  

At the Tevatron, experiment
E735 measured particle production at $p_T\sim 1\,$GeV in
Run~I \cite{Gutay}. The data can be interpreted as
showing some of the features expected with the onset
of quark-gluon deconfinement.

Such deconfinement is of course the domain of RHIC.
Recent results from STAR\cite{Oldenburg} on
gold-gold collisions show
large anisotropies in particle flow; this is expected
in a hydrodynamical picture of the collision
where there is an elliptical source
region (coming from the
overlap between two nuclei in a non-head-on
collsion). The behaviour of the velocity moments as $p_T$
increases is consistent with jet quenching.  More strikingly,
the clear back-to-back dijet topology that is seen in
$pp$ collisions at RHIC is absent in $Au\,Au$ collisions; 
the trigger jet 
remains, but with ``nothing'' recoiling against it. 
Qualitatively, this is just what is expected if a parton-parton
pair is produced near the edge of ``blob'' of quark-gluon
matter; a jet travelling outward will emerge unimpeded, while its
back-to-back companion has to travel through the deconfined
matter and is rescattered and quenched.  

\subsection*{Diffraction}

Presentations on diffraction at HERA \cite{Levy},
CDF \cite{Goulianos} and D\O\ \cite{Repond}
together with a theoretical view \cite{Forshaw} showed 
that there is still much to understand in this area. 

A rapidity gap
is a region of phase space with no particles flowing
into it; a gap is expected if the hard scattering
involves exchange of a colourless object like a pomeron.
In $p\overline p$ collisions, such a gap would 
often be spoiled
by particles from spectator parton interactions and so
the gap survival probability is expected to be small.
Even though factorization (the apparent partonic structure
of an exchanged pomeron) is not guaranteed to work in $p\overline p$
collisions, if one simply takes pomeron parton densities inferred
from HERA together with the guess that there is a
10\% rapidity gap survival probability, then one can
describe CDF and D\O\ data quite well.  
In this picture
the pomeron is about 80\% gluonic.  

D\O\ showed a new result on diffractive $W$ production
tagged with rapidity gaps
(complementing CDF's earlier result). 
$W$ production with a rapidity gap is observed at
perhaps 1\% of the inclusive $W$ rate.  If true, this
is certainly surprising: how can one kick a parton
out of a proton with $Q^2=m_W^2$ and not destroy the
proton in the process?  Moreover, how can one
do this fully 10\% of the
time that one makes a $W$? (assuming a 10\% gap
survival probability and a 1\% measured rate).
Does this tell us something about the makeup of the
proton -- or about the underlying event in $p\overline p$
collisions?

In Run~II, both CDF and D\O\ are improving their diffractive
instrumentation.  CDF have added shower counters and calorimeters
to cover $3.5<|\eta|<7.5$; D\O\ have new Roman Pots in the
$\pm z$ direction and veto counters covering  $2.5<|\eta|<6$.
Some Run~II physics goals that even the ``pomeron skeptical''
could support include making a direct measurement of the
rapidity gap survival probability, indeed testing the
assumption that rapidity gaps correlate with diffracted
(anti-)protons seen in the Roman Pots.  It will also be
good to measure the cross section for the process
$p \overline p \rightarrow p\,(gap)\, jj\, (gap)\,\overline p$.
This will provide a sanity check for ideas of Higgs production
through 
$pp \rightarrow p\,(gap)\,H\, (gap)\,p$ at the LHC. Published
cross sections for the latter process cover three orders of magnitude 
with prospects ranging from ``promising'' to ``impossible''.

\subsection{CKM Physics}

Our goal here is to confront the unitarity triangle
in ways that are complementary to the $e^+e^-$ 
$B$-factories\cite{Fleischer}.
CP violation is now established in the $B$ system through
$B_d \rightarrow J/\psi\, K^0_S$.  We find $\sin\phi_d = 0.734 \pm 0.054$;
either $\phi_d = 47^\circ$ ($=\sin 2\beta$ in the SM) or
$133^\circ$ (new physics).  BaBar and BELLE can and will
do much more with their data, for example, is $B\rightarrow
\pi K$ consistent with $\gamma < 90^\circ$ (SM)?  Is the
mixing asymmetry the same in $B_d \rightarrow J/\psi\, K^0_S$
and $B_d \rightarrow \phi K^0_S$?  $B\rightarrow \pi\pi$ will
be an important piece of the puzzle, and right now the
two experiments' measurements of this asymmetry 
are not really consistent. 

The ``El Dorado'' for hadron collider experiments (first at
the Tevatron,
then at the LHC\cite{Merk}) is the $B_s$ system. 
They will measure the mixing parameter $x_s=\Delta m_s/\Gamma_s$ 
and thus determine the length of the
unitarity-triangle side opposite the angle $\gamma$. 
CDF expect to have sensitivity to SM values of $x_s$ with a few hundred 
pb$^{-1}$, and to reach up to $x_s = 70$ with 2\,fb$^{-1}$.
The width difference $\Delta\Gamma_s/\Gamma_s$ may also
be significant in the $B_s$ system.  The angle $\gamma$
can be extracted using the decay $B_s \rightarrow D_s K$. 
It will be interesting to see if there is sizeable
CP violation in $B_s \rightarrow J/\psi\,\phi$ (it should
be small in the SM); also, $B_s \rightarrow KK$ complements
$B_d \rightarrow \pi\pi$ and together they can pin down
$\gamma$.  There are many other interesting topics such
as rare decays (e.g. $B \rightarrow K^* \mu^+\mu^-$,
$B_{s,d}\rightarrow \mu^+\mu^-$) and relating CP
violation in the $B$ system with $K\rightarrow \pi\nu\overline\nu$.

CDF reported\cite{Cecco} most impressive results
from Run~II, building on their Run~I experience together
with the new detector capabilities (silicon vertex trigger,
time of flight detector).  With a leptonic trigger, signals for
$B^+ \rightarrow J/\psi\, K^+$,
$B^0 \rightarrow J/\psi\, K^{*0}$, and
$B_s \rightarrow J/\psi\, \phi$ are seen; using the SVT,
the purely hadronic modes
$B^\pm \rightarrow D^0\pi \rightarrow K\pi\pi$ and 
$B \rightarrow {\rm hadron}\,{\rm hadron}$ are
being recorded.  We can also look forward to the
world's largest sample of charm mesons.

In D\O\, the tools are being put in place for a $B$-physics
program\cite{Taylor}. The inclusive $B$ lifetime
has been measured and D\O's first $B$ mesons are
being reconstructed ($B^\pm \rightarrow J/\psi K^\pm$).
D\O\ can not make use of purely hadronic triggers but benefits
from its large muon acceptance, forward tracking coverage,
and ability to exploit $J/\psi \rightarrow e^+e^-$.  

\subsection{Electroweak Physics}

Here we wish to indirectly probe new physics through
its virtual effects on precision electroweak observables. At
hadron colliders, the most powerful constraints come
from our measurements of the
masses of the $W$ boson and the top quark.  

\begin{figure}[t]
\begin{center}
\includegraphics[width=0.67\textwidth]{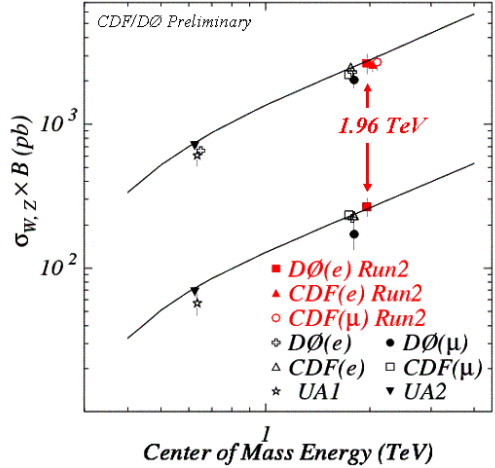}
\end{center}
\caption[]{Cross section times
leptonic branching ratio for $p\overline p\to W/Z \to \ell$.
The data points at
1.96~TeV are new, preliminary Run~II measurements}
\label{eps1}
\end{figure}

D\O\cite{Alton} and CDF\cite{James} both
reported first results from Run~II samples of $W$ and
$Z$ candidates. The experiments have measured the
cross section times branching ratio to leptons $\sigma\cdot B$ 
the new centre of mass energy of 1.96\,TeV (Fig.~\ref{eps1})
and also the
ratio of $\sigma_W\cdot B(W\to\ell\nu)/\sigma_Z\cdot B(Z\to\ell\ell)$ 
which allows an indirect extraction
of the $W$ width. CDF also have taken a first look at
the forward-backward asymmetry in $e^+e^-$ production
in Run~II.

Currently, the hadron collider determination of
$m_W$ is $80\,454\pm59$\,MeV, while the world average
(which is dominated by LEP) is $m_W = 80\,451\pm33$\,MeV.  
The high precision achieved at LEP means that 
it will take a hadron collider dataset of order 1\,fb$^{-1}$ in Run~II
before we can significantly tighten our knowledge of
$m_W$ -- this is not a short term prospect.  Given 15\,fb$^{-1}$,
however, we will eventually be able to drive the
uncertainty down to the 15\,MeV level.  Even greater
precision will be possible thanks to the huge statistics
at the LHC\cite{Pallin} should it be needed.

\subsection{The Top Quark}

We wish to measure the top quark's properties with greatly
increased statistics, and also use it as a possible
window to new physics.  Both D\O\cite{Johns} and CDF\cite{Wagner} 
showed ``steps on the road to
rediscovering top'' and both experiments have candidate
events.  In contrast to the case with the $W$ mass, we can
look forward to significant improvements in the short to
medium term because the Run~I dataset was so statistically
limited.  We expect of order 500 $b$-tagged $t\overline t$
events in the lepton $+$ jets final state, per fb$^{-1}$
recorded.  We plan to improve the cross section and
mass measurements, look
for $t\overline t$ spin correlations, and observe single top
production (which yields a model-independent measurement of the
CKM matrix element $|V_{tb}|$).

D\O\ also reported\cite{Estrada}
a significant improvement in the
extraction of the top mass using existing Run~I lepton $+$
jets data.  The new technique makes use of more information
per event: it involves calculating a likelihood as a function of
$m_t$, for both signal and background hypotheses, event by
event. All possible jet assignments and neutrino 
momenta are considered.  The event likelihoods are then combined 
to give an 
overall likelihood curve from which $m_t$ is extracted.
This technique gives better discrimination between
signal and background than the published 1998 analysis\cite{d0top}
and improves the statistical error equivalently to
a factor 2.4 increase in the number of events.  The new
result is
\begin{equation}
m_t = 179.9 \pm 3.6 {\rm (stat.)}\pm 
6.0{\rm(sys.)}\,{\rm GeV}\ \  {\it Preliminary}
\end{equation}
Use of the hadronic $W$ decay offers hope to reduce the
current systematic error, which is dominated by jet energy scale.

Some beyond-the-standard-model theories predict unusual
top properties and/or states decaying into top that would
be visible in Run~II or at the LHC\cite{Simmons}.  
Examples include
$X \rightarrow t\overline t$ (searched for by CDF and D\O,
who reported a new Run~I limit at this meeting), a top-Higgs
with flavour-changing decays visible as a $t+{\rm jet}$
resonance, and anomalously enhanced single top production.

\subsection{The Higgs Boson}

Our goal is to discover (or exclude) the SM Higgs
and/or the multiple Higgs bosons of supersymmetry. We
want to observe as many production modes and decay
channels as possible so that we can combine measurements
to extract the Higgs couplings.  

At the Tevatron\cite{Murat,Choi}, the emphasis
now is on developing the foundations needed for Higgs 
physics: good jet resolutions, $b$-tagging and trigger
efficiencies, and a good understanding of all the
backgrounds. One area that can be attacked with
relatively modest luminosities in Run~II
is associated production
of a neutral SUSY Higgs with a $b\overline b$ pair; at high
$\tan\beta$ the cross section is enhanced and Run~I
data have already allowed limits to be set.    
 
To explore the full range of SM Higgs masses will require
$10-15$\,fb$^{-1}$ and that, in turn, dictates upgrades
to the CDF and D\O\ detectors and trigger systems.  These
upgrades are now moving towards approval with installation
planned for 2005.

On the theoretical front \cite{Harlander} there are
improved predictions both for signals and backgrounds.
Backgrounds are available at NLO level using {\tt mcfm} and
DIPHOX programs.  The $t\overline t$ process, from which
the top quark Yukawa coupling can be extracted, is now
calculated at NLO: the Tevatron cross section is 
20\% lower than earlier (LO) estimates, while the LHC
is 20\% higher. 
The $gg\rightarrow H$ process is even available at NNLO
(resulting in much reduced scale-dependence) and the $p_T$ 
and $y$ distributions at NLO.
At the LHC\cite{Mellado},  there has been a lot of interest in the
Weak-boson fusion process for Higgs production as a
way of accessing $H\rightarrow \tau\tau$ and $b\overline b$
decays --  both
as a low mass Higgs discovery mode and in order to
make coupling measurements.

\subsection{Searches for Physics beyond the Standard Model}

In Run~I \cite{Kim}, CDF and D\O\ carried out extensive 
searches for supersymmetry:
squarks and gluinos through
$E_T^{\rm miss} + {\rm jets (+\,lepton(s))}$; charginos and neutralinos through
multileptons; gauge mediated SUSY through 
$E_T^{\rm miss} + {\rm photon(s)}$;
stop and sbottom; and $R$-parity violating variants.
Searches for other new phenomena included leptoquarks, dijet
resonances, $W^\prime$ and $Z^\prime$, 
massive stable particles,
monopoles, and extra dimensions.  In all cases, no new
physics was found. (At this meeting, CDF did report a possible
disagreement between the observed and predicted number of
lepton$+$photon$+E_T^{\rm miss}$ events. This may be 
something to watch at HCP 2004?)

D\O\cite{Nomerotski} has embarked on a number of searches 
using Run~II data.  
Work has started on understanding the $E_T^{\rm miss}$
distribution in multijet events as a prelude to squark
and gluino searches; trilepton candidates are also being
accumulated.  A gauge mediated SUSY search
has been carried through
to set a limit of 0.9\,pb on the cross section for
$p\overline p \to \gamma\gamma+E_T^{\rm miss}$.
Virtual effects of extra dimensions are being 
sought in $p\overline p\to e^+e^-,\ \mu^+\mu^-,\ {\rm and}\
\gamma\gamma$.  Limits of $M_S(GRW)>0.92(0.50)\,$TeV are
set in the electron/photon (and muon) final states.
Also, a search for leptoquarks decaying to electron$+$jet
excludes masses less than 113\,GeV (for $B({\rm LQ}\to ej)=1$).
None of these cross sections or mass limits is
better, yet, than published Run~I limits, but serves 
as a demonstration that the pieces are all in place.  

There is also a niche for searches at HERA.  The existing
data no longer indicate any high-$Q^2$ excess but there
are some deviations in leptonic final states\cite{Gallo} (H1 see an
excess of $e/\mu + E_T^{\rm miss}$, $ee$ and $eee$ events;
ZEUS do not, but have an excess of taus).  These are not
compatible with any obvious ``new physics'' explanation
and we will have to wait for HERA~II data\cite{Sirois}.  
HERA~II benefits
from significantly increased luminosity; its
polarization could be an important
tool to disentangle any eventual signal.  HERA complements
the searches that will be carried out at the Tevatron on
the same timescale, looking for $R$-parity violating
stop production, FCNC in the stop sector, doubly charged
Higgs particles, extra dimensions, leptoquarks, and
lepton flavour violation. 

The combination of 14\,TeV and 100\,fb$^{-1}$
makes the LHC an extremely potent discovery machine.
It is impossible to prove a theorem that any new physics
associated with the TeV scale will be detectable at LHC, but
``proof by enumeration'' has been carried a long way.
It will be hard for SUSY to escape detection\cite{Acosta}.  
The mass reach is
up to 2.5\,TeV for squarks and gluinos in minimal SUGRA,
and exclusive mass reconstruction has been demonstrated
at several benchmark points. Other new physics
detectable\cite{Chumney} includes extra dimensions and
TeV-scale gravity (the gamut from
indirect effects to black hole production); compositeness
(up to scales 20--40\,TeV); excited quarks, technicolor,
strong $WW$ scattering, leptoquarks, new gauge bosons,
and heavy right handed neutrinos.

\section{Our future}

It is impossible to avoid the feeling that hadron collider
physics has been wandering in the wilderness for the past
year. Whichever side
of the Atlantic we have been on, we seem to have been 
beset with problems:  technical, financial, management,
schedule, and politics.
We need to keep the faith --
we must remember that the physics remains the best in the world.
We also have a vibrant, enthusiastic community of young physicists.
There are clear indications that point to the existence of
physics beyond the
Standard Model:  neutrino mixing requires a new mass scale;
astrophysics and cosmology require dark matter and maybe stranger
things yet; 
the overall rather poor $\chi^2$ of global electroweak fits may be
telling us something; the questions of masses and
mixing angles 
remain unanswered; the equal electric charges of the electron
and proton surely require some kind of grand unification; and
the origin of the 246\,GeV weak scale is still unknown. 
Electroweak symmetry breaking and the Higgs is central to
all of these.  It is the key question for the Standard Model
and a window to beyond-the-Standard-Model physics.  

\begin{figure}[t]
\begin{center}
\includegraphics[width=\textwidth]{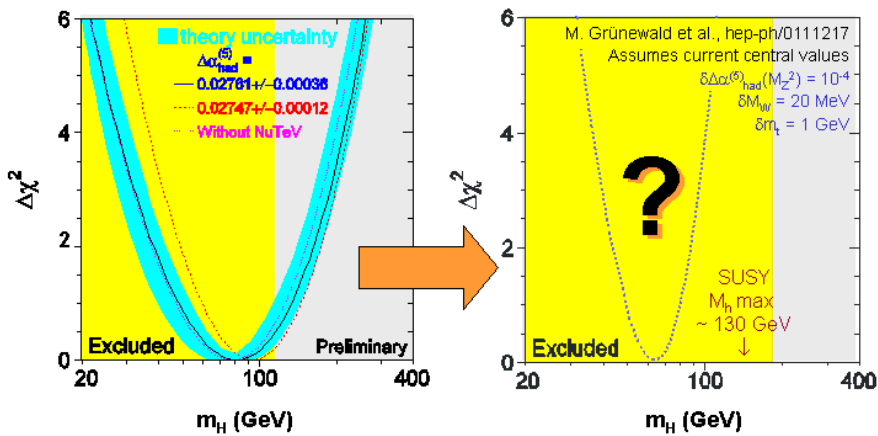}
\end{center}
\caption[]{Current (left) and anticipated post-Run~II 
(right)\protect\cite{snow} status of indirect and direct
constraints on the standard model Higgs.
Indirect constraints are shown by the parabolic curve and
direct exclusion by the yellow shaded region.
The current plot is from the LEP electroweak working group.}
\label{eps2}
\end{figure}

In the short term, we can look forward to physics results
from Run~II with few hundred inverse picobarns.  This is
a significantly increased sample over Run~I with improved
detectors and a higher centre of mass energy.
We can expect results on 
\begin{itemize}
\item a first look at $B^0_s$ mixing, 
\item top quark measurements with increased statistics and purity,
\item jet cross sections at high $E_T$ (constraining the gluon PDF),
\item new limits on physics beyond the SM (e.g. MSSM $(A/H)$ at
large $\tan\beta$).
\item ...
\end{itemize}
In the longer term, it will be a disappointment if Run~II does not
tell us something about electroweak symmetry breaking.  
The goal should be to transform both our indirect and direct
knowledge of the Higgs as shown in Fig.~\ref{eps2}\cite{snow}, 
or better yet to make this plot irrelevant.  This way we can lay
the foundations for a successful LHC physics program --
and hopefully a linear collider to follow.

%

\end{document}